\documentstyle[preprint,aps]{revtex}
\tightenlines
\begin{document}
\draft
\title{On the superluminal propagation of light-pulses in a
  "transparent" dispersive medium }
\author{Bruno Macke\thanks{Electronic address:
    bruno.macke@univ-lille1.fr} and Bernard S\'egard}
\address{Laboratoire de Physique des
Lasers, Atomes et Mol\'ecules,\\
Centre d'\'Etudes et de
Recherches Lasers et Applications,\\
Universit\'e de Lille 1, F-59655 Villeneuve d'Ascq (France)}
\date{\today}
\maketitle

\begin{abstract}
The mechanisms leading to a seemingly superluminal propagation of light in
dispersive media are examined. The anomalous dispersion near an absorption
line, involved in the first experiments displaying negative group
velocity propagation, is shown to be not only simpler but also more
efficient than the so-called transparent dispersion used in the later
experiments of Wang {\it et al}. (Nature (London) {\bf 406}, 277
(2000)) and of Dogariu {\it et al}. (Phys.Rev.A {\bf 63}, 053806
(2001)). The modest pulse advances observed by these authors can in
particular be reproduced with a better overall transmittance of the
medium and the significant advances attained with a single-line
arrangement would probably be unattainable with their gain-doublet arrangement.
\end{abstract}


\newpage

Much ado was made about the late experimental evidence of superluminal
propagation of light-pulses in an amplifying medium \cite{wang1,wang2}. The
repeated statement that ''the peak of the pulse exits the cell before it
even enters''\cite{wang1,wang2} is in part responsible for the hazardous
speculations on ''superluminality'' that appeared in the press. This
excitement should not however hide the fact that the pulse advances reported by Wang {\it et al}. \cite{wang1} and by
Dogariu {\it et al}. \cite{wang2} are modest (a few percent of the
pulse width). More significant advances have been observed a long time
ago by Chu and Wong \cite{chu3} and by S\'{e}gard and Macke
\cite{segard4} in experiments exploiting the very large anomalous
dispersion occurring near an absorption line. Astonishingly enough,
Dogariu {\it et al}. hastily discard these experiments under the
pretext that, due to the absorption, they ''resulted in controversies
over the interpretations'' \cite{wang1,wang2} and even that the phenomenon was
''poorly understood'' \cite{wang2} ! We are not aware of the controversies
alluded to by Dogariu {\it et al}. and everybody can appreciate the degree
of physical understanding of the authors of Ref.\cite{chu3} and Ref.
\cite{segard4} by examining the corresponding papers. As shown hereafter, the
physics underlying their experiments and those of Dogariu {\it et al}. is in
fact the same but the dispersion mechanism is more efficient in the
former that in the latter. Consequently the pulse advances observed by
Dogariu {\it et al}. in a medium which is globally absorbing
(transmittance 40\%) can be reproduced with a better transmittance and
the significant pulse advances obtained in Ref.\cite{chu3} and
Ref.\cite{segard4} would be probably impossible to attain with their
gain-doublet arrangement. 

Both kinds of experiments discussed here involve negative group velocities.
Garrett and McCumber \cite{garrett5} have been the first to recognise that,
under suitable conditions, pulses of light can actually propagate at such
velocities. A bell-shaped pulse then travels in the medium without
significant alteration in such a manner that the peak of the pulse exits the
medium while the signal is still increasing at its entrance. It is worth
emphasising that this does not imply any breaking of the causality
principle \cite{garrett5,diener6,chiao7} since ''the $t<0$ envelope maximum
seen by an observer at $z>0$ is not a direct reflection of the maximum of
the input-pulse envelope, but arises from the action of the dispersive
medium on the weak early components of that envelope''
\cite{garrett5}. From a spectral point of view, the phenomenon
exclusively results from an interference between the different
frequency components of the pulse in an anomalous dispersion region and not from a more or less selective amplification or absorption in
the medium, as stated in Ref.\cite{wang1}. Variations in the transmittance of
the medium with frequency have in fact detrimental effects : by
privileging some frequency components, they result in dramatic distortions
of the pulse and/or changes of its mean frequency, which increase rapidly
with propagation distance. Unfortunately, the large anomalous dispersions
required to attain significant effects are always associated with rapid
variations in the transmittance of the medium with frequency. The best
results are expected to be obtained when there exists a particular frequency
for which the transmittance and the (negative) group velocity are
simultaneously extremum, that is nearly constant in the vicinity of
this frequency. This condition is fulfilled in both kinds of
experiments underdiscussion \cite{wang1,wang2,chu3,segard4}. The
working frequencies are indeed close to a minimum of transmittance,
respectively a maximum of absorption \cite{chu3,segard4} or a minimum
of gain \cite{wang1,wang2}, coinciding with an extremum of the group velocity.

Explicit expressions of the advance of the pulse and of its distortion
are conveniently derived from the transfer function which links the
Fourier transforms of the complex envelopes of the input and output fields. By using the retarded time $t^{\prime}$
($t^{\prime}=t-z/c$ in a dilute sample), this transfer function may be
written $H(u)=\exp\left [G(u)\right ]$ where $G(u)$ is the complex gain and
$u=(\omega-\omega_{0})$ is the deviation of the frequency $\omega $
from the central frequency $\omega_{0}$ of the pulse spectrum, taken
as origin. When the anomalous dispersion originates from an ensemble
of $n$ homogeneously broadened lines, the complex gain simply reads :
\begin{equation}\label{equat1}
\Gamma(u)=\sum_{i=1}^{n}\left( \frac{G_{i}}{1+iu/\gamma +i\delta_{i}/\gamma_{i}}\right)
\end{equation}
with $\delta_{i}=(\omega_{0}-\omega_{i})$ and where $\omega_{i}$, $G_{i}$
and $\gamma_{i}$ are respectively the frequency, the amplitude gain and the
relaxation rate of (or corresponding to) the $i^{th}$ line.
$G_{i}=g_{i}L/2$, where $g_{i}$ is the intensity gain coefficient on
resonance (negative for an absorption line) and L is the medium
thickness. Eq.\ref{equat1} is obtained by assuming that the pulse
bandwidth and all $\delta_{i}$ and $\gamma_{i}$ are small with
respect to $\omega_{0}$ and that all $g_{i}$ are small compared to the
wavenumber $\omega_{0}/c$ (slowly varing amplitude and phase
approximation). Note that the poles (singularities) of $\Gamma(p)$ and
then of the transfer function $H(p)$ with $p=iu$ are located in the
left part of the complex plane. This ensures that the impulse response
$h(t^{\prime })=\int_{-\infty }^{+\infty }H\left(u\right)
\exp(iut^{\prime })du/2\pi$ of the medium is null for $t^{\prime }<0$
and, $t^{\prime }$ indicating here the retarded time, that
superluminal propagation of information  is forbidden
\cite{diener6,chiao7}. A general study of the systems described by
Eq. \ref{equat1} will be presented in a forthcoming paper. The analysis
is here restricted to the cases where the anomalous dispersion
originates either from a single absorption line as in the experiments
of Chu and Wong \cite{chu3} and of S\'{e}gard and Macke \cite{segard4}
or from a doublet of two identical amplifying lines as in the
experiments of Dogariu {\it et al}. \cite{wang1,wang2}. The detunings $\delta_{i}$ are chosen such that the real and imaginary parts of the
complex gain are respectively even and odd functions of
$u=(\omega-\omega_{0})$. This warrantees that the medium transmittance
and the group velocity will be both extremum for $\omega =\omega_{0}$
and is achieved for $\delta_{1}=0$ in the single-line case $(n=1)$ and
for $\delta_{1}=-\delta_{2}$ in the gain-doublet case $(n=2)$. The
complex gains, respectively $\Gamma_{1}$ and $\Gamma_{2}$, take then the simple forms given below.
\begin{equation}\label{equat2}
\Gamma_{1}(u)=-\frac{Z}{1+iu/\gamma }
\end{equation}
where Z is the optical thickness of the medium ($Z=\alpha
L/2=-G_{1}>0$, $\alpha$ being the intensity absorption coefficient on resonance).
\begin{equation}\label{equat3}
\Gamma_{2}(u)=\frac{G}{1+iu/\gamma -i\Delta }+\frac{G}{1+iu/\gamma +i\Delta }
\end{equation}
where $G_{1}=G_{2}=G>0$ and $\Delta =-\delta_{1}/\gamma =+\delta_{2}/\gamma >0$.
If the pulse spectrum is very narrow with rapidly decreasing wings,
the complex gain may be approximated by its first order expansion in ($u/\gamma $).
The transfer functions $H_{1}(u)$ and $H_{2}(u)$ can then be written
under the general form $H_{n}(u)=H_{n}(0)\exp\left(iua_{n}\right)$ with
\begin{equation}\label{equat4}
a_{1}=Z/\gamma>0
\end{equation}
\begin{equation}\label{equat5}
H_{1}(0)\ =\exp\left (-Z\right)= \exp\left (-\gamma a_{1}\right)<1
\end{equation}
\begin{equation}\label{equat6}
a_{2}=\frac{2G}{\gamma }\cdot \frac{\Delta ^{2}-1}{\left( \Delta
^{2}+1\right) ^{2}}
\end{equation}
\begin{equation}\label{equat7}
  H_{2}(0)=\exp\left (\frac{2G}{1+\Delta ^{2}}\right)= \exp\left (\gamma a_{2}\frac{\Delta ^{2}+1}{\Delta ^{2}-1}\right) >1
\end{equation}
$H_{n}(0)$ and $a_{n}$ being real \cite{com8} for $n = 1$ or 2, the effect
of the medium on the complex envelope of the pulse is to change its
amplitude by the factor $H_{n}(0)$ and to shift it in time by $a_{n}$.
The time-shift is easily identified with the advance of the pulse
having propagated in the medium at the group velocity on the same
pulse having covered the same distance in vacuum. $a_{2}$ is positive
(true advance) only if $\Delta>1$ and attains its maximum value
$G/4\gamma$ for $\Delta=\sqrt{3}$. As indicated before, the amplitude
transmittance of the medium $\left|H_{n}(u)\right|$ is minimum for
$u=0$. It is an absolute minimum for $\left| H_{1}(u)\right|$ and a
relative minimum between two maximums for $\left| H_{2}(u)\right|$
when $a_{2}>0$ (see Fig.5 of Ref.\cite{wang2}). The essential point is
that the larger are the aimed advances the more the transmittances
depart from unity, downwards in the single-line case and upwards in
the gain-doublet case (see Eq.\ref{equat5} and Eq.\ref{equat7}). It is thus
impossible to obtain advances without change of the pulse amplitude,
unless one associates amplification and absorption. This has been simply
achieved in the experiments \cite{wang1,wang2,chu3,segard4} by
compensating the absorption (resp. the amplification) by a suitable
broadband, i.e. non-dispersive, amplification (resp. absorption). The
device achieving the required amplification (resp. absorption) should
be considered as an integral part of the systems under investigation
(denoted $S_{1}$ and $S_{2}$ in the following). Absorption and
amplification appear then to be tightly coupled in both kinds of
experiments which, from this point of view, only differ by the
implementation of the dispersion, associated either to the absorption stage \cite{chu3,segard4} or to the amplification stage \cite{wang1,wang2}.

The pulse advances should naturally be referred to the pulse duration
$\tau_{p}$ (hereafter the half-width at $1/e$ of the amplitude
profile) and the relative merits of the systems $S_{1}$ and
$S_{2}$ are suitably evaluated by comparing the pulse distortion
they generate for the same value of the relative advance $A=a/\tau
_{p}$, where $a$ indicates $a_{1}$ or $a_{2}$. This distortion
can be derived from the exact form of the transfer functions of the
above-defined global systems $S_{1}$ and $S_{2}$. The transfer
function $H_{1}^{\prime }$ of $S_{1}$ can be expressed in terms of the
dimensionless quantities $A$, $Z$ and $\Omega= (\omega -\omega_{0})\tau _{p}$. One gets:
\begin{equation}\label{equat8}
H_{1}^{\prime }(\Omega )=\exp \left( \frac{i\Omega A}{1+i\Omega A/Z}\right) 
\end{equation}
with $A=Z/\gamma \tau_{p}$. The equivalent expression of the transfer
function $H_{2}^{\prime }(\Omega )$ of $S_{2}$ is somewhat complicated
in the general case but only its particular form for $\Delta
=\sqrt{3}$ (leading to the maximum advance) is needed for the
following discussion. It reads : 
\begin{equation}\label{equat9}
H_{2}^{\prime }(\Omega )=\exp \left( i\Omega A\frac{1-2i\Omega A/G}{
1+2i\Omega A/G-4\Omega ^{2}A^{2}/G^{2}}\right)
\end{equation}
where $A=G/4\gamma \tau_{p}$. Note that $H_{1}^{\prime}(\Omega )$ and
$H_{2}^{\prime }(\Omega )$ both tend towards a constant when $\Omega
\rightarrow \infty $\ . The corresponding impulse response
$h_{1}^{\prime }(\theta )$ and $h_{2}^{\prime }(\theta )$ with
$\theta=t^{\prime }/\tau_{p}$ are then proportional to the Dirac
function $\delta(\theta )$ and this means that the propagation of
temporally localised features (involving high frequency components) is
luminal. Advances are obtained in the opposite limit $\Omega
\rightarrow 0$ and the distortion of the pulse envelope, if it is not
too large, mainly originates from the second order term in the
expansion of the complex gain in powers of $\Omega$. The transfer
functions $H_{1}^{\prime }(\Omega )$ and $H_{2}^{\prime}(\Omega )$ can then be written under the common form.
\begin{equation}\label{equat10}
H_{n}^{\prime}\left( \Omega \right) \approx \left( 1+\varepsilon_{n} \Omega ^{2}/2\right)\exp \left( i\Omega A\right)
\end{equation}
with $\varepsilon_{1}=2A^{2}/Z$ and $\varepsilon_{2}=8A^{2}/G$. The
transcription of Eq.\ref{equat10} in the time domain \cite{macke9} leads to :
\begin{equation}\label{equat11}
E(L,\theta )\approx E(0,\theta +A)-\frac{\varepsilon }{2}\frac{d^{2}}{d\theta ^{2}}E(0,\theta +A)
\end{equation}
where $E(L,\theta )$ and $E(0,\theta )$ are respectively the complex
envelopes of the output and input pulses and where $\varepsilon$
indicates $\varepsilon_{1}$ or $\varepsilon_{2}$. The last term
of Eq.\ref{equat11} is obviously responsible for the pulse distortion
which, for a given pulse shape, is fully characterised by the
parameter $\varepsilon$. Although superluminal propagation is not
specific to gaussian pulses \cite{macke9}, these pulses are suitable
for simple calculations \cite{wang1,wang2,chu3,segard4,garrett5,macke9}. Putting $E(0,\theta )=\exp (-\theta ^{2})$ in Eq.\ref{equat11}, it is easily
shown that the alteration of the pulse simply consists of a magnification of
the amplitude profile by the factor $\beta=(1+\varepsilon )$ and its
narrowing by the same factor, in agreement with the area theorem for the
field envelope \cite{crisp10}. The intensity profile is also narrowed
by the factor $\beta $ but magnified by the factor $\beta ^{2}$. The
narrowing is in particular responsible for the difference between the
advances $a_{\downarrow }$ and $a_{\uparrow }$ on the fall and the rise of
the pulse. Such a difference is particularly apparent on Fig.6 of Ref.\cite{wang2}. Dogariu {\it et al}. claim that ''the front edges and the trailing
edges of the pulse are shifted forward by the same amount''. A simple
inspection of the insets in their figure shows on the contrary that
$a_{\downarrow }$ and $a_{\uparrow }$ are dramatically different, roughly in
the proportion of 3 to 2. Quite generally,
$a_{\downarrow}=a+\tau_{a}(1-\frac{1}{\beta })$ and
$a_{\uparrow}=a-\tau_{a}(1-\frac{1}{\beta })$, where $\tau_{a}$
indicates the half-width of the pulse at the amplitude or
intensity used to determine $a_{\downarrow }$ and $a_{\uparrow }$. The
difference ($a_{\downarrow }$ - $a_{\uparrow }$) is suitably
characterised by the ''contrast'' $\xi $ :
\begin{equation}\label{equat12}
\xi =\frac{a_{\downarrow }-a_{\uparrow }}{a_{\downarrow }+a_{\uparrow }}
\approx \frac{a_{\downarrow }-a_{\uparrow }}{2a}\approx \frac{\varepsilon
\tau _{a}}{a}
\end{equation}
The parameter $\xi $ has the advantage of referring the distortion to the relative advance. 

Since the pulse advances observed with the gain-doublet arrangement
\cite{wang1,wang2} are one order of magnitude smaller than those
attained with the single-line scheme \cite{chu3,segard4}, the direct
comparison of the absolute distortions is meaningless. These
distortions indeed increase as the square of the advances. It is
obviously more relevant to examine the potentialities of each
experiment for a same pulse advance and a same level of distortion
($A$ and $\varepsilon$ given) or, more precisely, to determine in what
conditions each system would be able to reproduce the results obtained
with the other. This is easily achieved by remarking that, $A$ being
fixed, the same level of distortion is attained in both arrangements
when $Z=G/4$ (see the expressions of $\varepsilon $).

In the experiment of Dogariu {\it et al}. \cite{wang2}, $G$, $a$ and the
full width at half maximum of the intensity profile of the pulse are
respectively 0.7, 63 ns and 2.4 $\mu s$. This leads to
$\tau_{p}= 2.04 \mu s$ (half-width at $1/e$ of the corresponding
amplitude profile, assumed to be gaussian), $A=0.031$ and
$\varepsilon = 0.011$. The value $\Delta =\sqrt{3}$, used to establish
the relation $\varepsilon =8A^{2}/G$, is consistent with the contrast
$\xi$ deduced from Fig.6 of Ref.\cite{wang2} ($\xi\approx 0.2$) and
with the separation of the two maximums of gain shown in Fig.5. The
same relative advance and distortion would be obtained in the
single-line case with an optical thickness $Z=G/4$ and the intensity
transmittance of the medium would be $\exp(-G/2)\approx 0.7$. It is amazing to note that the
transmittance of this ''absorbing'' medium is better than that of the
''transparent'' medium used by Dogariu {\it et al}. (only 0.4, due to
the presence of reverse-pumped atoms).

On an other hand one may wonder whether it would be possible to
observe the large advances attained in the single-line case
\cite{chu3,segard4} with a gain-doublet arrangement. For the sake of
definiteness, the data of the Fig.2 of Ref.\cite{segard4} are
considered. $Z$, $a$ and the half-width at half-maximum of the
amplitude profile of the pulse are respectively 5.25, 2 $\mu s $ and 4
$\mu s$. This leads to $\tau_{p}= 4.8 \mu s $, $A= 0.42$, $\varepsilon
=0.067$ and $\xi=0.13$. The extended system $S_{1}$ under consideration consists of the medium of amplitude transmittance
$e^{-Z}$ and an amplifier of amplification factor $e^{Z}$ ($\approx
$46dB). Assuming again $\Delta =\sqrt{3}$, the same relative advance
and distortion would be obtained in the gain-doublet case with
$G=4Z=21$. The maximum of $Re(\Gamma_{2})$ is $G/(4\sqrt{3}-6)\approx
1.08 G$ whereas its minimum (at $\Omega=0$) is $G/2$. The maximum of
the field amplification factor of the global system $S_{2}$ would
then be $e^{0.58G}\approx2.10^{5}$ $(106dB)$. Even if it would be
feasible (?) a system having such an optical gain at some frequency would be unstable. It is obviously possible to reduce the
optical gain by a broadband absorber overcompensating the gain at
$\Omega =0$ and to use a post-amplifier (as it is actually done in the
experiments of Dogariu {\it et al}.). However the
amplification factor of the required post-amplifier would be
considerably larger than that of the amplifier used in the single-line
arrangement and this could raise serious problem of signal to noise ratio.

In conclusion, the propagation of light pulses at a negative
group velocity in a medium whose dispersion originates from a single
absorption line is a phenomenon which has been well understood for a
long time and now pertains to classical physics \cite{jack11}. The
physics underlying the experiments where the dispersion results
from a doublet of two amplifying lines is quite similar. Indeed both
kinds of experiments involve tightly coupled amplification and
absorption and, in principle, only differ by the stage where the
dispersion is implemented. However, the single-line arrangement
appears, in practice, to be more efficient that the gain-doublet
scheme. A study of other arrangements is in progress. It
appears possible in particular to cancel the second order distortion
(prevailing in the experiments discussed above) by using a doublet of
absorption lines. In all cases, obtaining significant advances with
moderate distortion requires the association of very large
amplification and absorption. Otherwise said, Mother Nature
resists to any transgression of her laws, even when this trangression
is only apparent as in the experiments of superluminal propagation.

The authors thank J. C. Garreau and A. Walters for a critical reading of
the manuscript. The Laboratoire de Physique des Atomes, Lasers et
Mol\'{e}cules is ''Unit\'{e} Mixte de Recherche de l'Universit\'{e} de
Lille 1 et du CNRS (UMR 8523)''. The Centre d'Etudes et de Recherches
Lasers et Applications (CERLA) is supported by the Minist\`{e}re charg\'{e} de
la Recherche, the R\'{e}gion Nord-Pas de Calais and the Fonds
Europ\'{e}en de D\'{e}veloppement Economique des R\'{e}gions.

{\it \addcontentsline{toc}{chapter}{Bibliographie} }


\begin{references}
\bibitem{com8} If $a_{n}$ is complex (as it occurs for detunings $\delta_i$
different from those considered here), its imaginary part results in a shift
of the mean frequency of the pulse and generally a distortion (see, e.g.,
the next reference).
\end{references}
\end{document}